\documentclass[useAMS,usenatbib,usegraphicx]{mn2e}
\usepackage{amsmath,mathtools,amsfonts,amssymb,float}
\usepackage[usenames,dvipsnames]{xcolor}
\usepackage{pdfpages}
\usepackage{soul}
\usepackage{subfig} 
\usepackage{graphicx}
\usepackage{calc}

\newcommand{\twoc}{_{\textrm{200c}}}

\def\Msun{\hbox{$\rm\, M_{\odot}$}}

\title [mass scales for star formation and quenching]{The origin of the mass scales for maximal star formation efficiency and quenching: the critical role of Supernovae}

\author[Bruno M. B. Henriques et al.]  
{Bruno M. B. Henriques$^{1}$\thanks{E-mail:brunohe@phys.ethz.ch},
Simon D. M. White$^{2}$, Simon J. Lilly$^{1}$, 
\newauthor 
Eric F. Bell$^{3}$,  Asa F. L. Bluck$^{1}$, Bryan A. Terrazas$^{3}$\vspace{0.4cm}\\
{}$^{1}$Department of Physics, ETH Zurich, CH-8093 Zurich, Switzerland\\
{}$^{2}$Max-Planck-Institut f\"ur Astrophysik, Karl-Schwarzschild-Str. 1, D-85741 Garching b. M\"unchen, Germany\\ 
{}$^{3}$Department of Astronomy, University of Michigan, Ann Arbor, MI 48109, USA\\
}
 
\begin{document}

\date{Submitted to MNRAS, 2018}

\volume{000}
\pagerange{0000--0000} \pubyear{2018}

\maketitle

\label{firstpage}

\begin{abstract}
We use the \citet{Henriques2015} version of the Munich galaxy
formation model ({\small L-GALAXIES}) to investigate why the halo and stellar mass scales
above which galaxies are quenched are constant with redshift and
coincide with the scale where baryons are most efficiently converted
into stars. This model assumes that central galaxies are quenched by
AGN feedback when hot halo gas accretes onto a supermassive black
hole. Nevertheless, we find that supernova (SN) feedback sets both
mass scales. As haloes grow above a threshold mass, SNe can no longer
eject material so their hot gas content increases, enhancing the
cooling rate onto the central galaxy, its cold gas content, its star
formation rate and the growth rate of its central black hole. Strong
AGN feedback terminates this short-lived phase by suppressing the fuel
supply for star formation. Despite strong evolution of the halo mass -- temperature 
relation, quenching occurs at a redshift-independent halo
and stellar mass which coincides with the mass where baryons have been
converted into stars with maximal efficiency. These regularities and
coincidences are a result of the specific parameters selected by MCMC
tuning of the model to fit the observed abundance and passive fraction
of galaxies over the redshift range $0\leq z\leq 3$. Thus they are required by
the observed evolution of the galaxy population, at least in the
context of models of this type.

 \end{abstract}


\section{Introduction}
\label{sec:intro}
One of the most striking features of the low-redshift galaxy
population is that more than half of the stellar mass density is in
quenched galaxies \citep{Kauffmann2003a, Baldry2004, Bell2004, Bundy2005,
  Faber2007}. This value drops below $50\%$ at $z=1$ and to $\sim25\%$
at $z=2$ \citep{Ilbert2013, Muzzin2013, Tomczak2013}.  The likelihood
of a galaxy being quenched has been found to correlate with a number
of other galaxy (e.g., stellar mass, halo mass, black hole mass, central density)
and environmental (e.g, central vs. satellite, halo mass, group centric radius) properties. 
Stellar mass has received particular attention, with massive
galaxies found more likely to be quenched \citep{Kauffmann2003b,
  Brinchmann2004, Baldry2004} and to be quenched earlier
\citep{Thomas2005, Bundy2005, Faber2007, Scarlata2007, Ilbert2013, Muzzin2013,
  Tomczak2013}.  \citet{Peng2010} showed that both trends are
reproduced if the probability of a quenching event is assumed to
increase strongly with stellar mass in a redshift-independent way (at
least for $z\lesssim1$).

Quenching the most massive galaxies requires some dramatic
phenomena. Since the early work of \citet{White1991} it has been clear
that gas cooling rates in massive haloes will result in overly massive
central galaxies unless they are substantially offset by some heating
process; the development of static hot atmospheres and the reduced
halo accretion rates in dark energy dominated cosmologies are
insufficient on their own to explain why such a small fraction of the
baryons associated with massive haloes lie in their central
galaxies. Successful models for the formation of the galaxy population
in a $\Lambda$CDM universe are forced to include an additional
feedback mechanism to suppress cooling of gas in such haloes, with AGN
feedback being the preferred candidate \citep{Springel2005b, Croton2006,
  Bower2006}. However, it remains unclear why feedback from hot gas
accretion onto a central black hole results in such a well defined and
redshift-independent stellar mass at quenching.

Another striking feature of the galaxy population is that the ratio
between the stellar mass of a central galaxy and the mass of its halo
shows a sharp peak at $\log_{10} (M_{\rm 200c}/\Msun) \sim 12.0$, 
 corresponding to a maximum efficiency of $\sim20\%$ for
converting baryons into stars \citep{Guo2010, Moster2010}. This
efficiency falls off quite strongly for halo masses above and below
this characteristic value which again appears to be almost independent
of redshift \citep{Moster2012, Behroozi2013}. As highlighted by
\citet{Birrer2014}, and more recently by \citet{Puebla2017}, this scale
of maximal efficiency seems to coincide with the mass above which
galaxies are typically quenched. This coincidence and the invariance
of this mass scale, at least since $z\sim3$, seems particularly
puzzling given the strong evolution in the typical density and
temperature of halos of given mass.

In the \citet{Henriques2015} model, AGN feedback suppresses 
cooling and star formation in massive galaxies, resulting in a 
strong dependence between quenching and black hole mass as observed
 \citep{Bluck2016, Terrazas2016a, Terrazas2016b, Terrazas2017},
although the transition may be too abrupt in the model. In addition, 
\citet{Henriques2016} showed that this particular implementation of AGN
feedback couples with environmental effects to reproduce quite well the 
trends with stellar mass and environment identified by \citet{Peng2010}.    
Here we investigate {\it why}
the interplay between the different feedback mechanisms in our 
model also reproduces the remarkably strong observed dependence of
quenching and star formation efficiency on halo/stellar 
mass and the relative weak dependence of these relations on redshift. 
Of particular relevance is the emergence of this behaviour in
a model where star formation strangulation from AGN feedback is a 
consequence of black hole growth without requiring a direct dependance with 
stellar or halo mass. 

\section{The model}
\label{sec:munich_model}

The model used here is fully described in the supplementary material
of \citet{Henriques2015}\footnote{The supplementary material is
  attached to the main paper at arXiv:1410.0365. Snapshot and
  light-cone catalogues are publicly available at:
  http://www.mpa-garching.mpg.de/millennium. The source code which
  generated the model is now also public at:
  http://galformod.mpa-garching.mpg.de/public/LGalaxies/. This page
  also contains catalogues in fits format and additional material
  related to the model.} (hereafter H15).  The evolution of the
baryonic components -- including hot, cold and ejected gas, stars in
disks, bulges and haloes, and black holes -- is self-consistently
coupled to the evolution of dark matter subhaloes identified in
dark-matter-only N-body simulations. These are scaled to the {\it
  Planck} cosmology ($\sigma_8=0.829$,
$H_0=67.3\;\rm{km}\;\rm{s}^{-1}\rm{Mpc}^{-1}$,
$\Omega_{\Lambda}=0.685$, $\Omega_{\rm{m}}=0.315$,
$\Omega_{\rm{b}}=0.0487$, $f_{\rm{b}}=0.155$ and $n=0.96$) according
to the procedures of \citet{Angulo2014}. Specifically we use the
Millennium \citep{Springel2005a} and Millennium-II \citep{Boylan2009}
simulations which each trace $2160^3$ ($\sim$10 billion) particles
from z = 127 to the present day. The Millennium was carried out in a
box of original side $500\,h^{-1}\rm{Mpc}=685\,\rm{Mpc}$.  After
rescaling to the {\it Planck} cosmology, this becomes 714~Mpc,
implying a particle mass of $1.43\times10^{9}\Msun$. The Millennium-II
follows a region one fifth the linear size, resulting in 125 times
better mass resolution. Combined, the two simulations follow dark
matter haloes which host galaxies spanning five orders of magnitude in
stellar mass at $z=0$. Throughout the paper, the Millennium-II
Simulation is used for $\log_{10} (M_*/\Msun) \leq 10.0$ and 
$\log_{10}(M_{200c}/\Msun) \leq 11$ and the Millennium Simulation for 
higher stellar masses. Above this mass cut, properties of galaxies are 
nearly identical in the two simulations. The exception are 
properties that strongly depend on merger histories for which we 
restrict our analysis to MRII (this is the case in Fig.\ref{fig:channels}).

The self-consistent treatment of the evolution of the stellar, gas and
black hole components of galaxies assumes that baryons and dark matter
are initially fully mixed and that collapsed dark matter structures
have a baryon fraction given by the cosmic mean for the adopted
cosmology. Baryons are initially assumed to be in the form of diffuse
hot gas which either cools immediately or forms a static hot
atmosphere with a central cooling flow. Cold gas settles in a disk and
fuels star formation which eventually leads to the release of energy
and metals through SN. This can either reheat gas back into the hot
halo phase or eject it entirely into an external reservoir from which it
might be subsequently reincorporated. Mergers result in starbursts and 
drive the growth of both bulges and central black holes through cold gas 
accretion. Subsequent hot gas accretion onto black holes generates 
AGN ``radio-mode" feedback which can completely shut down cooling. 
Whenever galaxies become satellites, and before they merge with their 
central companions, environmental effects gradually remove their hot, 
cold and stellar components.

Of particular relevance for the present study is our adopted implementation 
of SN feedback. Following \citet{Guo2011} the efficiencies of reheating 
(of cold gas into the hot phase) and ejection (of hot gas into the external 
reservoir) are assumed to scale inversely with the virial velocity of the 
halo, while the timescale for reincorporation of ejected gas scales 
inversely with virial mass \citep{Henriques2013}. Also important are our black 
hole growth and AGN feedback models. Black hole growth is dominated by 
cold gas accretion during mergers which is assumed to be more efficient in 
more massive systems \citep{Kauffmann2000} while their feedback is
determined by hot gas accretion. The energy output is assumed to scale
as the product of the black hole and hot gas masses, and is used to
offset the cooling of hot gas (\citealt{Croton2006}; H15).

In order to better understand the impact of different feedback channels 
in the conversion of baryons into stars and in star formation quenching,
throughout the paper we will analyse the results from four variations of 
the H15 model: the original model, a model with no AGN and no SN 
feedback (no-feedback model), a model without SN feedback (AGN-only 
model) and a model without AGN feedback (SN-only model). For all model 
variations the parameters are kept constant at the H15 values, set by the MCMC 
sampling constrained by  the observed abundance and passive fraction of galaxies 
over the redshift range $0\leq z\leq 3$, which makes it easier to isolate the impact 
of different feedback channels.

\begin{figure}
\centering
\includegraphics[width=1.0\linewidth]{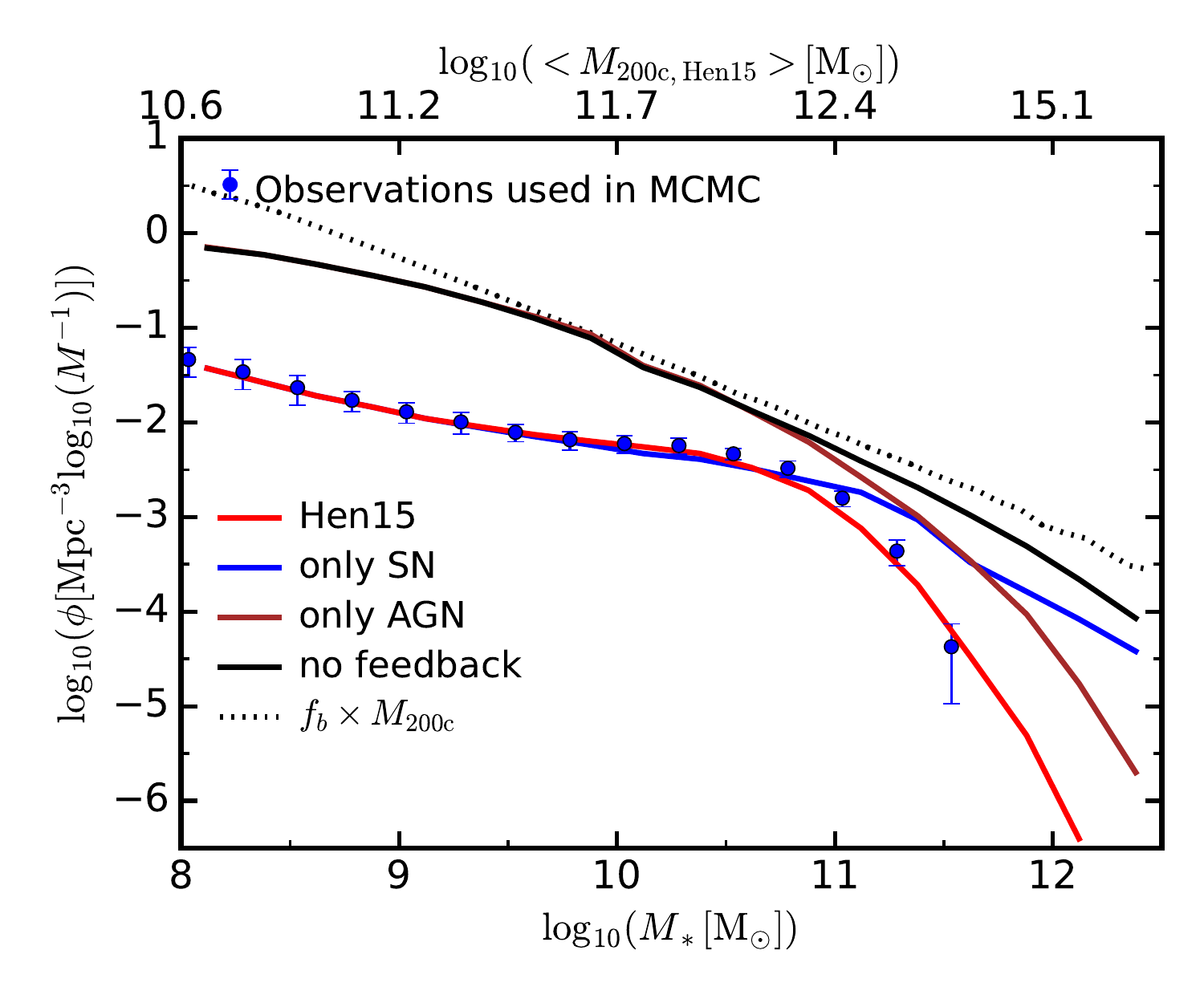}
\caption{Stellar mass functions at $z=0$ for different variations
  of the H15 model (shown as a solid red line). The solid black,
  blue and brown lines show, respectively, versions of the H15 model
  with neither AGN nor SN feedback, with SN feedback only (no AGN) and
  with AGN feedback only (no SN). The dotted black line shows the halo
  mass function scaled by the baryon fraction. The halo mass scale on the upper
  $x$-axis applies to the full H15 model only. The Millennium-II
Simulation is used for $\log_{10} (M_*/\Msun) \leq 10.0$ and the Millennium 
Simulation for higher stellar masses. }
\label{fig:smf}
\end{figure}

\begin{figure*}
\centering
\includegraphics[width=1.0\linewidth]{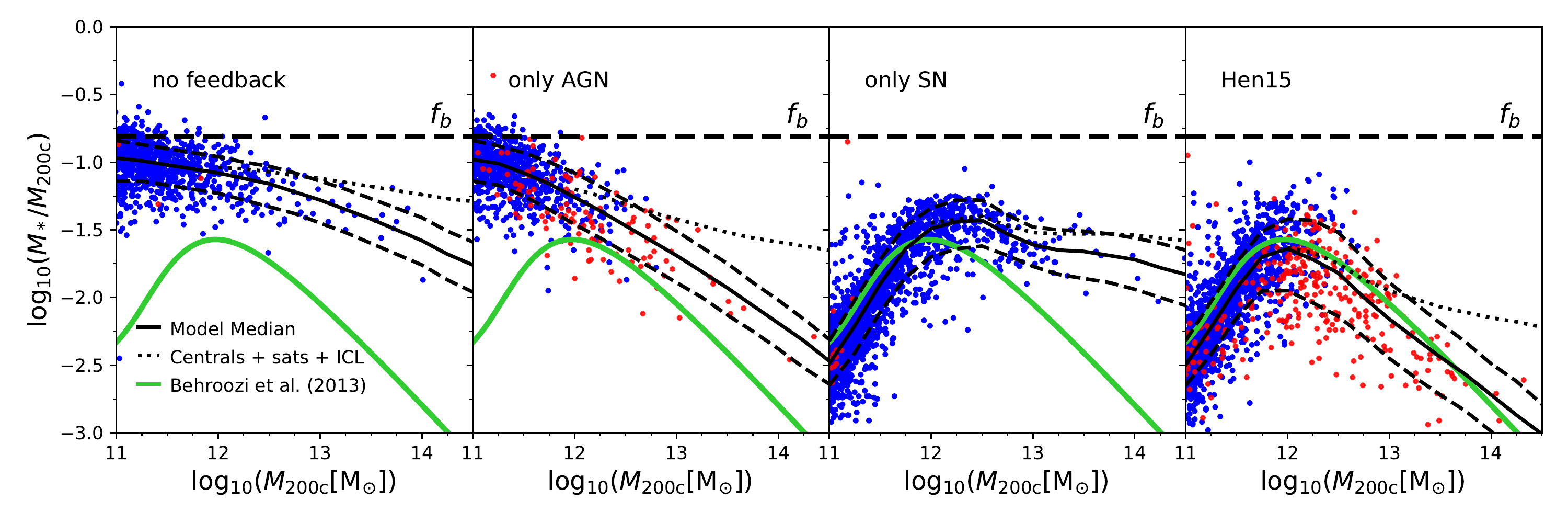}
\caption{The stellar-to-halo mass ratio versus halo mass for central galaxies in
the four models analysed in the present study: no feedback (left panel), AGN
  feedback only (middle-left panel), SN feedback only (middle-right
  panel) and the standard H15 model (right panel). The coloured dots
  show a random sample of galaxies in each model, with blue being
  star-forming and red being passive objects (divided at
  $\log_{10}(sSFR[\rm{yr}^{-1}])=-11.0$). The solid and dashed black
  lines show, respectively, the median and $1\sigma$ scatter around
  the median for the different models, while the solid green line shows
  an abundance matching estimate from \citet{Behroozi2013}. 
The dotted black line shows the model median including satellites and  ICL. A clear peak
  in the distribution appears when SN feedback is included
  (middle-right panel) which coincides with the mass scales at which
  galaxies transition from star-forming to quenched when AGN feedback
  is present (right panel).}
\label{fig:smhm}
\end{figure*}

\section{Results}

In our self-consistent model for the evolution of galaxies, energy
feedback from SN and AGN competes with accretion, cooling and star
formation to establish the rates for conversion of baryons into cold
gas and stars.  We will start our analysis by trying to understand how
these different mechanisms combine to produce, at $z=0$, a
characteristic stellar mass above which galaxies tend to be
quenched. We also investigate why this quenching threshold coincides
with the halo mass at which the integrated efficiency of converting
baryons into stars is maximal. We will then investigate why this
characteristic mass scale remains constant out to at least $z\sim3$.

\subsection{The stellar mass function at $z=0$: a mass scale for peak efficiency of converting baryons to stars}
\label{subsec:smf}

In Fig.~\ref{fig:smf} we show stellar mass functions at z=0 for the
four models used in our study: the standard H15 model (solid red
line), the no-feedback model (solid black line), the AGN-only model
(solid brown line) and the SN-only model (Solid blue line). In
addition, the black dotted line shows the halo mass function
multiplied by the cosmic fraction of baryons in our cosmology
(0.155). The top x-axis shows the average $M_{\rm{200c}}$ for a given
stellar mass in the H15 model (note that this does {\it not}
correspond to the mean halo mass at this stellar mass in the other
three models). We see that without any form of feedback (solid black
line) most of the baryons are converted into stars and the stellar
mass function closely follows the re-scaled halo mass function (dotted
black line). The exceptions are the very low- and high-mass ends 
where, respectively, photo-heating by the UV background and the development
of a hot atmosphere make cooling less efficient. In addition, at high-mass
a significant fraction of the baryons are in satellite galaxies and the
intracluster light.

When AGN feedback alone is included (solid brown line) the high-mass
tail of the stellar mass function is significantly reduced above
$\log_{10}(M_*/\Msun) \sim11$, corresponding in this model to
$\log_{10}(M_{\rm{200c}}/\Msun) \sim 12.2$, while the low-mass end
remains unchanged. The onset of AGN feedback and the suppression
of cooling and galaxy growth happens at the observed stellar mass scale, 
coinciding with the formation of a substantial hot halo. This is a 
consequence of the assumed relation between the strength of radio mode 
feedback and hot gas mass. However, as we will see in section~\ref{sec:smf_evo}, 
without SN winds this mass scale evolves too strongly with redshift. 
In addition, the reduction in the stellar content of galaxies inhabiting 
the most massive haloes is much less pronounced than in observations.

More dramatic changes arise when SN feedback is
included (solid blue line): there is a large reduction in the stellar
mass function at all masses with respect to the no-feedback case,
but particularly for low-mass galaxies.  A clear
characteristic mass appears, again at about $\log_{10}(M_*/\Msun)
\sim11$ corresponding in this case to $\log_{10}(M_{\rm{200c}}/\Msun)
\sim 12.4$, below which the reduction is more significant. The
reduction at high masses with respect to the no-feedback case is caused
both by the reheating of cold gas in massive galaxies 
(despite a failure to eject any hot gas) and by the reduced stellar 
masses of accreted satellites.

Once both feedback channels are included (solid red line) suppression is
effective at both low and high masses with the knee (where the stellar
mass function is closest to the scaled halo mass function) being
shifted to slightly lower mass than in the SN-only case.  This first
plot clearly shows that, while SN feedback is solely responsible for
the reduction in the number of low-mass galaxies, both SN and AGN feedback 
are required to explain the sharp cutoff at the high-mass end.
The combination of the two effects results in a relatively small
range in stellar mass for which the fraction of baryons converted into stars
is maximal.

\subsection{The stellar mass/halo mass relation at $z=0$: a quenching threshold 
that coincides with maximal star-formation efficiency}
\label{subsec:hmsm}

The stellar mass functions shown in the previous subsection seem to
indicate that the interplay between SN and AGN feedback plays a critical 
role in reducing the range in stellar  mass for which the fraction of baryons 
converted into stars is close to maximal and in establishing a sharp high-mass 
cut-off. Here we investigate this further by looking at the relation between
stellar-to-halo mass ratio and halo mass as a function of star formation 
activity in the central galaxy. This will help us to understand why the scale of 
maximal integrated star formation efficiency coincides with the threshold for 
quenching.

Fig.~\ref{fig:smhm} plots the ratio of central galaxy stellar mass to
halo mass against halo mass for our four models: no-feedback (left
panel), AGN-only (middle-left panel), SN-only (middle-right panel) and
the standard H15 model (right panel). Without feedback almost all
baryons are turned into stars in low-mass halos. At high mass, the
formation of a static hot halo delays cooling and reduces star formation. 
In addition, in this mass range, a significant fraction of the baryons
are in satellite galaxies or in the intracluster light (as shown by the dotted 
black line). However, as seen for the stellar mass function, this is insufficient 
to explain the very large reduction in baryon conversion efficiency in massive
haloes derived from abundance matching \citep[e.g.][the solid green 
line]{Behroozi2013}. It is also unable to explain why massive galaxies
are predominately quenched since these continue to form stars 
despite the delayed cooling.

When AGN feedback alone is included (second panel from the left) the
results are similar at low halo mass, but the star-formation efficiency is much 
lower at high mass. As explained in Section~\ref{subsec:heating_cooling}, 
radio-mode feedback is proportional to black hole and hot gas mass in the 
H15 model. Quenched galaxies appear and are spread relatively evenly 
(by number) across halo mass, and they predominate at high halo mass 
($\log_{10}  (M_{200c}/\Msun)>12.5$) as in the full model. However, 
without SN feedback, galaxies are extremely gas-rich, and large black 
holes and overly massive stellar populations can form even in low-mass
haloes. Large reservoirs of cold gas survive because AGN feedback is
not fueled by, and has no effect on cold gas in the H15 model. 
This results in a relatively small fraction of quenched galaxies
in lower mass haloes.

In contrast, for the SN-only case (second panel from the right), a
peak appears in the stellar-to-halo mass ratio. In the context of our model, 
at low masses the ejection 
of hot gas into the external reservoir greatly suppresses star formation. 
This ceases to be effective at $\log_{10}(M_{\rm{200c}}/\Msun)\sim12$ 
where the potential wells become too deep for ejection to be possible. 
At higher mass the conversion efficiency is still suppressed with respect to
the no-feedback case because cooling from the hot atmosphere is
partially offset by the return of disk gas reheated by SN and also because 
accreting satellites have fewer stars. Since SN regulate the supply of cold 
gas but are unable to eliminate it completely, there are no quenched central 
galaxies in this case. In our model, this only happens as a result
of AGN radio mode feedback from massive black holes. As detailed
below, SN are critical in establishing the halo/stellar mass scale at which 
black holes can grow and subsequently release energy, but are not
directly responsible for the quenching due to their inability to stop the hot gas 
atmosphere from cooling.

\begin{figure*}
\centering
\includegraphics[width=1.0\linewidth]{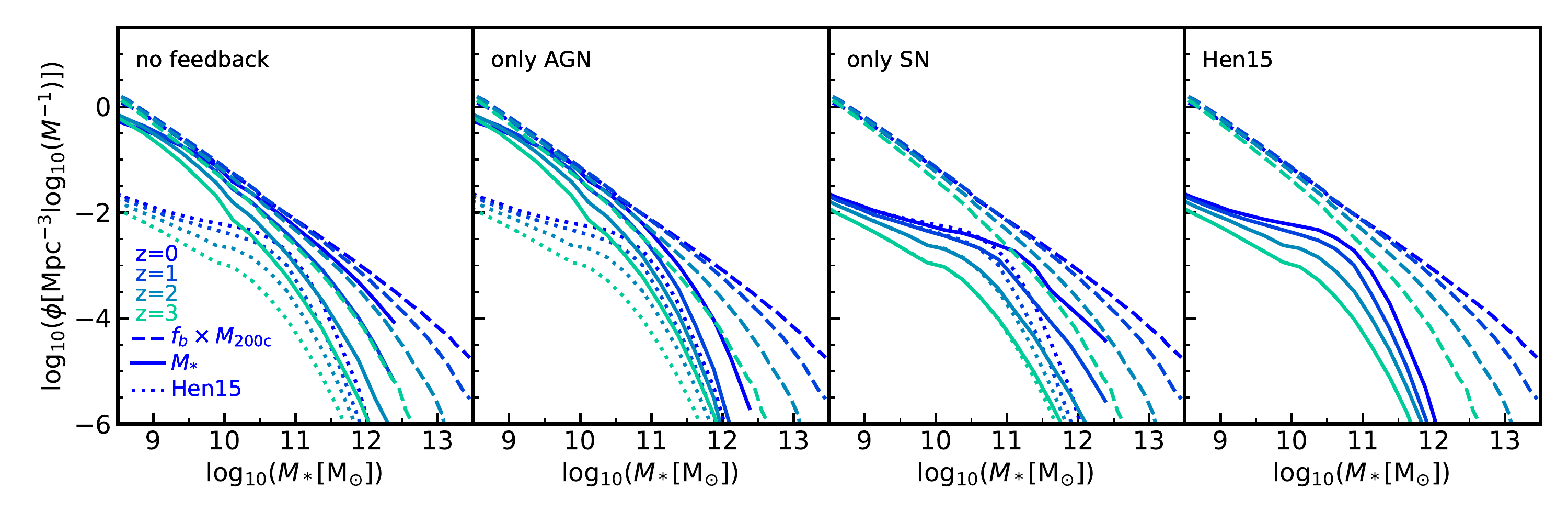}
\caption{The evolution from $z=3$ to $z=0$ of the stellar mass
  function (solid coloured lines) and of the halo mass function
  multiplied by the cosmic fraction of baryons (dashed coloured
  lines). From left to right, the different panels represent the
  no-feedback model, the AGN-only model, the SN-only model and the
  standard H15 model. In the three left panels, the coloured dotted
  lines show the evolution of the stellar mass function in the
  standard H15 model. The Millennium-II Simulation is used for 
  $\log_{10} (M_*/\Msun) \leq 10.0$ and the Millennium Simulation for higher 
  stellar masses (this explains the small discontinuity visible in the transition
between the two).}
\label{fig:smf_evo}
\end{figure*}

In the full H15 model where both SN and AGN feedback are included
(right panel), the peak in conversion efficiency created by SN
feedback remains. However, a clear threshold for quenching by AGN
feedback emerges at the same mass. The fraction of baryons in cold gas is 
relatively low in low-mass galaxies (with respect to their total baryon content)
and their assembly histories typically involve 
fewer of the equal-mass galaxy mergers required for black hole growth in the 
H15 model.\footnote{Note that \citet{Bell2017} argue that observationally it is unclear 
which mergers are associated with black hole growth.}
As a result, the ratio of black hole to stellar mass remains much lower than 
in more massive systems. Furthermore, efficient cooling leaves rather little
hot gas, so AGN feedback is negligible.  At the halo mass where
ejection ceases to be effective, the hot halos become more massive and
cooling from them causes the central galaxies to grow large cold gas
reservoirs. As will become clearer in the following sections, this
leads not only to maximally efficient star formation but also to increased
black hole growth. Shortly thereafter, the black holes become massive
enough for AGN feedback to suppress cooling altogether, thus
suppressing star formation. This is the apparent coincidence 
highlighted in the data by \citet{Birrer2014}.

In the H15 model, the relation between the efficiency of hot gas ejection by SN 
and $M_{\rm{200c}}$ is controlled by parameters which the MCMC sampling 
sets primarily in order to fit the observed stellar mass function at low mass. 
The detailed dependence results in SN being unable to eject gas from haloes 
more massive than $\log_{10}(M_{\rm{200c}}/\Msun)\sim12.$

\subsection{The evolution of the stellar mass function: a constant scale for peak efficiency of converting baryons to stars}
\label{sec:smf_evo}

\citet{Henriques2016} already gave indications that the stellar mass
threshold for quenching and the stellar mass at which the integrated
efficiency of star formation peaks are similar, and that both are constant with
redshift. The constancy of the former was suggested by the weak
evolution of the ``knee'' of the stellar mass function for
star-forming galaxies, while that of the latter was suggested by the
similarly constant knee in the stellar mass function for all
galaxies. Here we look in more detail at the evolution of the total
stellar mass function in order to better understand why the stellar
mass at the knee is constant despite the fact that the characteristics
of the corresponding haloes are evolving strongly.

In Fig.~\ref{fig:smf_evo} we plot the $z=3$ to $z=0$ evolution of the
stellar mass function for all galaxies as solid coloured lines, and
that of the halo mass function multiplied by the cosmic baryon
fraction as dashed coloured lines. From left to right, the different
panels represent the no-feedback model, the AGN-only model, the
SN-only model and the standard H15 model. In the three left panels,
the coloured dotted lines reproduce the stellar mass functions from
the standard H15 model. We see that, except for the full H15 model
(right panel), all stellar mass functions appear to show an evolving 
characteristic mass.  Without any feedback (left panel), the stellar 
mass functions at low mass correspond to almost 100\% conversion 
into stars of the available baryons. At higher mass the conversion is 
less efficient and the solid curves fall progressively to the left of the 
dashed ones. With decreasing redshift this reduction below 
100\% efficiency occurs at larger and larger stellar (and hence halo) mass. 

The introduction of AGN feedback (second panel from the left) has very little 
effect on the stellar mass functions at low mass ($ \log_{10} (M_*/M_\odot) < 10.5$) or high
redshift ($z\sim 3$) but at lower redshift the stellar masses of the
central galaxies in high-mass haloes are reduced, shifting the solid
curves further to the left. The evolution of the characteristic mass is reduced, 
although not as much as in the H15 model (and not to the observed level). 
In contrast, in the SN-only case (second panel from the right) the stellar mass 
functions are reduced substantially and deviations from the full H15 model 
are seen only at high mass and low redshift. At $z=3$ the stellar mass 
function of the SN-only model is essentially identical to that of the full model. 
Nevertheless, the characteristic mass evolves as in the no-feedback model. In the
 final H15 model (right panel), AGN feedback suppresses cooling 
 and star formation in  the most massive halos at later times. This effectively 
 eliminates the  progression of the knee of the stellar mass function towards 
 higher masses  which is seen in the SN-only model, resulting in an almost 
 constant characteristic mass, as required by the observations 
 ($ \log_{10} (M_*/M_\odot) \sim 11.4$, 10.9, 11.1 and 10.7, respectively 
 at $z=3, 2, 1$ and 0, see Fig.~2 of \citealt{Henriques2016}). In summary, 
 AGN and SN feedback alone can create a characteristic mass, this remains constant
at $z\le3$ when both are combined. In this case, as we will see in the next section, 
black holes can only grow from cold gas accretion above a very weakly evolving 
halo mass scale at which SN winds become inefficient.

\section{The origin of the characteristic mass scale}
\label{sec:physics}

Previous sections identified SN feedback as a necessary ingredient in
establishing a relatively narrow and weakly evolving mass scale 
at which the fraction of baryons converted into stars maximizes.
Although AGN quenching reduces the evolution of the characteristic mass of 
the stellar mass function, it only suppresses it completely when combined with SN. 
In addition, when SN are present, AGN produce a relatively sharp transition 
between haloes hosting star-forming and passive galaxies, that coincides 
with maximal efficiency of  baryon conversion into stars. In this section
we will investigate how individual physical processes interact to
produce this behaviour. 

\begin{figure}
\centering
\includegraphics[width=1.0\linewidth]{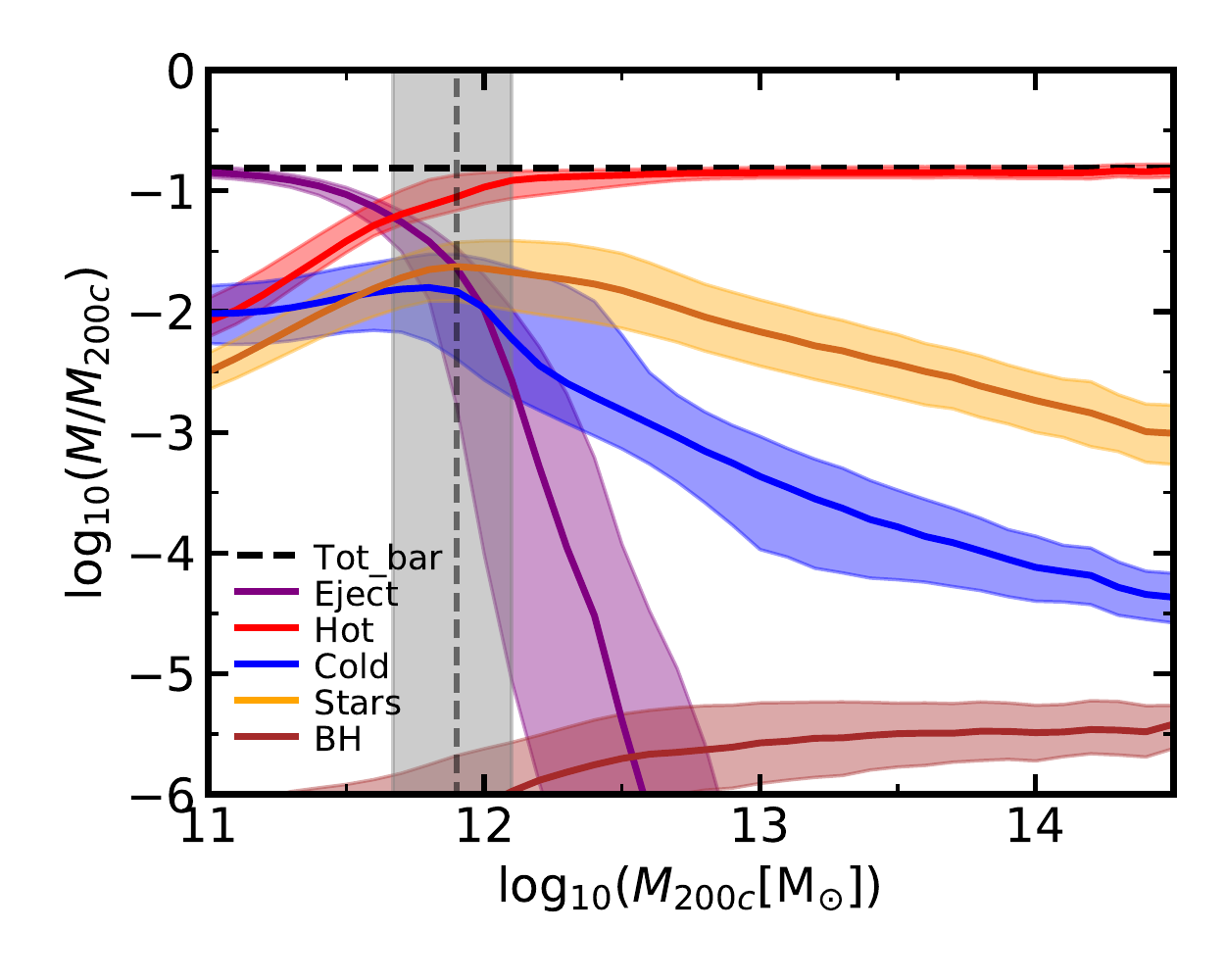}
\caption{Mass fraction in different components with respect to the
  virial mass as a function of the latter for central galaxies. The solid lines
  represent the median and the coloured regions the 16th to 84th
  percentiles of the distributions for
  $M_{\mathrm{eject}}/M_{\mathrm{200c}}$ (purple),
  $M_{\mathrm{hot}}/M_{\mathrm{200c}}$ (red),
  $M_{\mathrm{cold}}/M_{\mathrm{200c}}$ (blue), $M_*/M_{\mathrm{200c}}$
  (orange) and $M_{\mathrm{BH}}/M_{\mathrm{vir}}$ (brown). The dashed
  black line represents the median
  $M_{\mathrm{baryon}}/M_{\mathrm{200c}}$. The grey shaded region shows
  the halo mass between when the hot component becomes more massive than the external 
  reservoir (when reincorporation balances SN ejection) and when star formation is 
  quenched. The vertical dashed line marks the peak of star formation activity.}
\label{fig:mass_frac}
\end{figure}

\begin{figure*}
\centering
\includegraphics[width=1.0\linewidth]{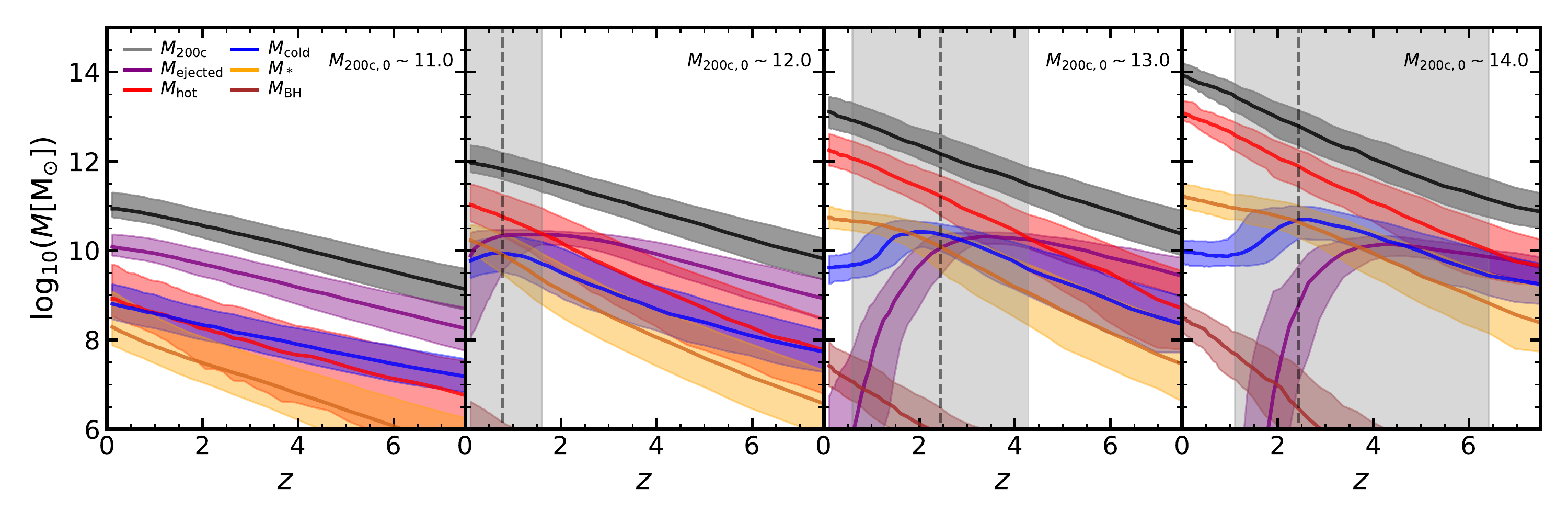}
\caption{The evolution of the mass in different components in the H15 model 
for objects which at $z=0$ lie at the centre of haloes with virial mass 
$10^{11}\Msun$ (left panel), $10^{12}\Msun$ (centre left panel), $10^{13}\Msun$ 
(centre right panel) and   $10^{14}\Msun$ (right panel). The solid lines 
represent the median   and the coloured regions the 16th to 84th percentiles 
of the distribution for $M_{\mathrm{200c}}$ (black), $M_{\mathrm{eject}}$
  (purple), $M_{\mathrm{hot}}$ (red), $M_{\mathrm{cold}}$ (blue),  $M_*$ (orange) 
  and $M_{\mathrm{BH}}$ (brown). The grey shaded region in the three right panels 
  highlights the redshift range between the time when the hot component becomes 
  more massive than the external reservoir and that when star formation is 
  quenched. The vertical dashed lines mark the peak of star formation activity.
 The Millennium-II Simulation is used in all panels except the right-most in order 
 to correctly follow the very low mass progenitors of the haloes considered.}
\label{fig:mass_frac_evo}
\end{figure*}

For reasons illustrated in
Figures~\ref{fig:mass_frac} and~\ref{fig:mass_frac_evo}, we find it
useful to distinguish three different halo mass regimes, corresponding
roughly to $\log_{10}(M_\mathrm{200c}/\Msun) < 12 $, 
$\log_{10}(M_\mathrm{200c}/ \Msun) \sim12$ and 
$\log_{10}(M_\mathrm{200c}/ \Msun )> 12$.  These figures
plot the median and 16 to 84\% scatter in the fractions of the total
baryons associated with five different components: the stars, the cold gas and
the supermassive black hole associated with the central galaxy, together with the
hot gas associated with the surrounding main subhalo, and the material
stored in the reservoir of halo ejecta (this represents material ejected
by SN winds, which might be reincorporated at later times into the hot gas). 
Fig.~\ref{fig:mass_frac} shows these fractions as a function of halo mass at $z=0$, while
Fig.~\ref{fig:mass_frac_evo} shows them in the form of the masses of
the individual components (and also of the halo itself) as a function
of redshift for halos which at $z=0$ have masses 
$\log_{10}(M_\mathrm{200c}/\Msun) \sim 11$, $12$, $13$ and $14$.
In both figures the grey shaded regions mark the transition phase between 
 when the hot component becomes more massive than the external 
  reservoir (when reincorporation balances SN ejection) and when star formation is 
  quenched ($sSFR<(1+z)/2t_{\mathrm{H}}$ is used to select passive 
  galaxies in the full model at different redshifts following Fig.~9 of \citealt{Terrazas2016a}), 
  while the vertical dashed lines mark the peak of star formation activity.

For low-mass haloes where cooling times are short, infalling material
shocks, cools and condenses as it hits the galaxy, thus immediately
becoming available to form stars. Since the SN ejection of hot gas 
into the external reservoir scales
inversely with $V_{\mathrm{max}}$ and gas reincorporation timescales
are long, most newly accreted gas is ejected from the halo by SN winds
and does not return. The baryons associated with such haloes are
therefore predominantly in the external reservoir, as can be seen in the 
purple line in Fig.~\ref{fig:mass_frac} at 
$\log_{10}(M_\mathrm{200c}/\Msun) < 12$. As $\log_{10}(M_\mathrm{200c}/\Msun)$
approaches $\sim 12$, SN cease to be effective at expelling gas, and 
the baryon fraction in the external reservoir drops dramatically (low mass 
boundary of grey shaded region). At the same time, hot gas cooling times 
lengthen to the point that a substantial hot component can be maintained, 
and infalling gas adds directly to this atmosphere (red line). However, 
cooling is still significant, leading to a peak both in the cold gas content 
(blue line) and in the specific star-formation rate (vertical dashed line). 
Since our model assumes central black holes to be built up by 
merger-driven accretion of cold gas, they grow rapidly in this regime 
(brown line) and soon reach masses for which AGN feedback provides 
enough energy to halt further cooling from the hot gas halo (high mass 
boundary of grey shaded region). At higher mass, $z=0$ halos have little cold 
gas, no external reservoir, and a declining baryon fraction in the central
stellar component. Interestingly, the model predicts central black hole 
mass to be proportional to halo mass rather than to central galaxy stellar
mass in this regime (cf \citealt{Bogdan2017}).

Fig.~\ref{fig:mass_frac_evo} shows that at all redshifts, low-mass
halos are in the same regime as at $z=0$ (left panel), while high-mass halos
enter the transition region between the two asymptotic regimes at the redshift 
when their progenitors have $\log_{10}(M_\mathrm{200c}/\Msun)$ just below 
12 ($z\sim 6$, $z\sim 4$ and $z\sim 2$, respectively in the first, second and 
third panels from the right).

\subsection{The SN feedback dominated and intermediate regimes: $M_\mathrm{200c} \le 10^{12} \Msun$}
\label{subsec:heating_cooling}

The behaviour of our model in the low-mass regime can be derived simply
from our underlying assumptions about cooling and the effects of SN
feedback. These are set out in detail in the supplementary material
(SM) to \citet{Henriques2015} which can be found online attached to
the arXiv preprint. The haloes of low-mass galaxies are always in the
rapid accretion regime where the bulk of newly accreted baryons are
assumed to fall onto the central galaxy and become available for star
formation on a short time-scale (see Section S1.4 of the SM). Because
such haloes have shallow potential wells, most newly accreted material
is immediately ejected from the halo in a SN-driven wind and the
amount of material turned into new stars is just that needed to
provide the requisite ejection energy. 

The energy input into outflowing gas is assumed to be directly
proportional to the mass of stars formed,
\begin{equation}
\Delta E_{\mathrm{SN}} = \epsilon_{\mathrm{halo}} \times  \frac{1}{2}\Delta M_*V^2_{\mathrm{SN}},
\end{equation}
where $V_{\mathrm{SN}} = 630$~km/s for the IMF we assume, and the
efficiency $\epsilon_{\mathrm{halo}}$ decreases with increasing $V_{200c}$, 
is unity for haloes with $V_{200c}\sim100$ km/s and drops to about 0.4 for 
large $V_{200c}$ (equation S17 of the SM). Some of this energy is used to 
reheat a mass:
\begin{equation}
\Delta M_{\mathrm{reheat}} = \epsilon_{\mathrm{disk}} \Delta M_*
\end{equation}
from the cold ISM to the hot halo, while the rest is used to drive gas
out of the hot halo into the external reservoir. The mass loading factor
$\epsilon_{\mathrm{disk}}$ varies with disk rotation velocity as
$8\times (V_{\rm rot}/100~{\rm km/s})^{-0.72}$ over the $V_{\rm rot}$
range of relevance (equation S19 of the SM). A specific energy
$\frac{1}{2}V^2_{200c}$ per unit mass is assumed to be necessary for
both the cold gas reheating and hot gas ejection processes resulting in:

\begin{equation}
\Delta M_{\mathrm{eject}}= (\epsilon_{\mathrm{halo}} V^2_{\mathrm{SN}}/V^2_{200c} -
\epsilon_{\mathrm{disk}}) \Delta M_* .
\label{eq:SN1}
\end{equation}
The material ejected from the hot halo is subsequently 
reincorporated at a rate given by:
\begin{equation}
\dot{M}_{\mathrm{ejected}}= -\frac{M_{\mathrm{eject}}} {t_{\mathrm{reinc}}},
\label{eq:reinc1}
\end{equation}
where $t_{\mathrm{reinc}}=\gamma 10^{10}/M_{200\mathrm{c}}$ and $\gamma=3\times10^{10} \rm{yr}$.

Assuming that the baryonic mass newly accreted onto the halo is 
$f_b \Delta M_{200c}$, where $f_b$ is the cosmic baryon fraction, that the 
sum of the mass changes in the cold and hot gas components is small 
enough to be neglected and that reincorporation times are long enough
to prevent any significant amount of material to return from the external
reservoir, we can write:
\begin{equation}
f_b \Delta M_{200c} \approx \Delta M_{\mathrm{eject}} + \Delta M_*.
\label{eq:SN2}
\end{equation}
This is the case for small haloes where the first term in the parenthesis on the
{\it rhs} of equation~\ref{eq:SN1} is also dominant and
$\epsilon_{\mathrm{halo}}=1$, giving:
\begin{equation}
f_b \Delta M_{200c} \approx \Delta M_{\mathrm{eject}} \approx (V^2_{\mathrm{SN}}/V^2_{200c}) \Delta M_*,
\label{eq:SN3}
\end{equation}
which, if haloes build up the bulk of their mass at approximately
constant $V_{200c}$, implies
\begin{equation}
M_*/M_{200c} \propto V^2_{200c} \propto (GM_{200c}H(z))^{2/3},
\label{eq:SN4}
\end{equation}
where $H(z)$ is the Hubble constant as a function of redshift. This
behaviour can be seen at low masses in Fig.~\ref{fig:mass_frac} (orange line).
It is also visible by comparing the four panels of Fig.~\ref{fig:mass_frac_evo} 
where both the stellar mass and the redshift of the point where the median 
progenitor halo mass is $10^{11}\Msun$ increase with increasing final halo 
mass (the value of the orange line when the black line reaches $10^{11}\Msun$).

As the virial velocity $V_{200c}$ of halos increases, the value of the
parenthesis on the {\it rhs} of equation~\ref{eq:SN1} drops, reaching
zero at $V_{200c}\sim 200$~km/s for the specific parameters of our
model. This corresponds to $\log_{10}(M_{200c}/\Msun)\sim 11.6$, 
11.8, 12.0 and 12.3 at $z=3$, 2, 1 and 0, respectively. In addition, the 
term in the bottom of equation~\ref{eq:reinc1} decreases and a significant 
amount of mass starts returning from the external reservoir. The 
reincorporation time becomes smaller than $0.1 \times t_{\mathrm{dyn}}$ 
for halo masses larger than 
$\log_{10}(M_{200c}/\Msun)\sim12.2$, 11.9, 11.7 and 11.5
at $z=3$, 2, 1 and 0, respectively, meaning that any ejected material 
is almost immediately reincorporated. The combined effect of reduced 
ejection into, and increased reincorporation from, the external 
reservoir {\it creates a characteristic mass,} $M'_{200c}$, {\it above 
which SN are unable to maintain gas outside haloes}. This mass is 
almost independent of redshift  (compare the high redshift boundary 
of the grey shaded regions in the three right panels of 
Fig.~\ref{fig:mass_frac_evo}, just below $\log_{10}(M_\mathrm{200c}/\Msun) \sim 12$).

Shortly after haloes reach this weakly evolving mass scale, 
there is a precipitous drop in the baryon fraction in the external reservoir and a 
corresponding increase in the amount of hot gas, which is the dominant 
baryonic component at $z=0$ in halos with 
$M_{200c}> 5\times 10^{11}\Msun$ (see purple and red lines in
Fig.~\ref{fig:mass_frac}, the increased fraction of hot gas is not as clear in 
Fig.~\ref{fig:mass_frac_evo} due to the many orders of magnitude plotted 
in the y-axis). The three right panels of Fig.~\ref{fig:mass_frac_evo} show 
that this transition occurs at $z\sim 2$,  $z\sim 4$ and $z\sim 6$ for the 
main progenitors of haloes with present-day mass 
$\log_{10}(M_{200c}/\Msun)\sim12$, $13$ and $14$, respectively. As 
expected, this corresponds to characteristic masses at that time which are 
similar to the present-day transition value. 
 
In our model, this mass roughly corresponds to the virial temperature at which 
haloes switch between the rapid cooling and cooling flow regimes. Again, 
this is a consequence of the particular parameter values picked out by our 
MCMC chains and is required in order for a substantial fraction of the
baryons to be able to remain in the hot phase for an extended
period. Nevertheless, cooling rates are still quite large at the
transition point, resulting in peaks in cold gas fraction and specific
star formation rate soon after haloes pass through this transition 
(vertical dashed lines in the three right panels of Fig.~\ref{fig:mass_frac_evo}). 
This causes an acceleration in black hole growth, and shortly thereafter 
the cold gas fraction drops dramatically as the majority of galaxies 
quench (the low redshift boundary of the grey shaded regions at $z\sim1.0$ and 
$z\sim0.5$ in the $\log_{10}(M_{\rm{200c}/\Msun})\sim$ 14 and 13 cases, 
respectively the first and second panels from the right in Fig.~\ref{fig:mass_frac_evo}).
Although cooling times lengthen in these more massive haloes as they grow,
these sharp drops in cold gas fractions require an additional quenching process. 
In our model this is AGN feedback, as we discuss next.

\begin{figure*}
\centering
\subfloat[AGN-feedback-only]{\includegraphics[width=.5\linewidth]{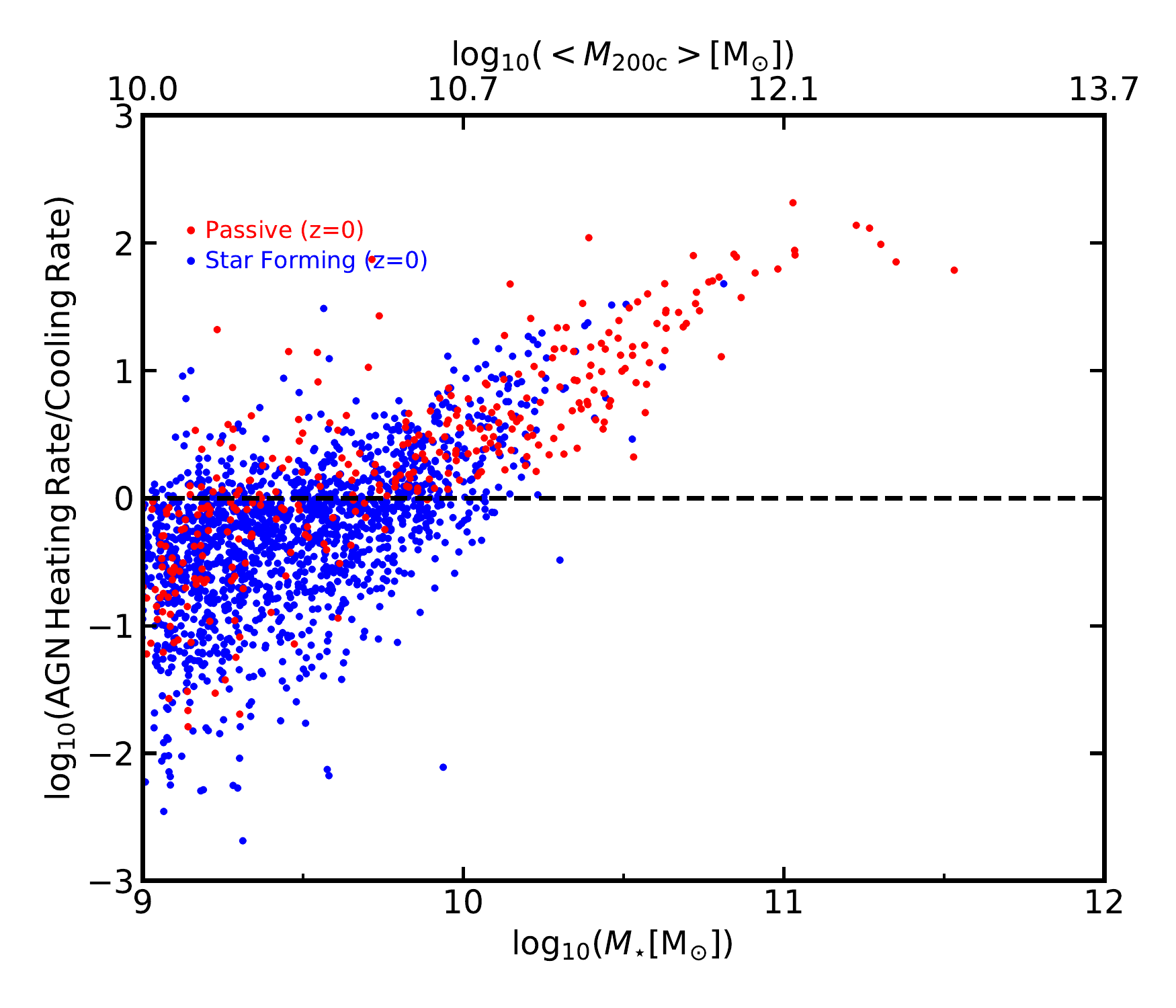}}
\subfloat[\citet{Henriques2015}]{\includegraphics[width=.5\linewidth]{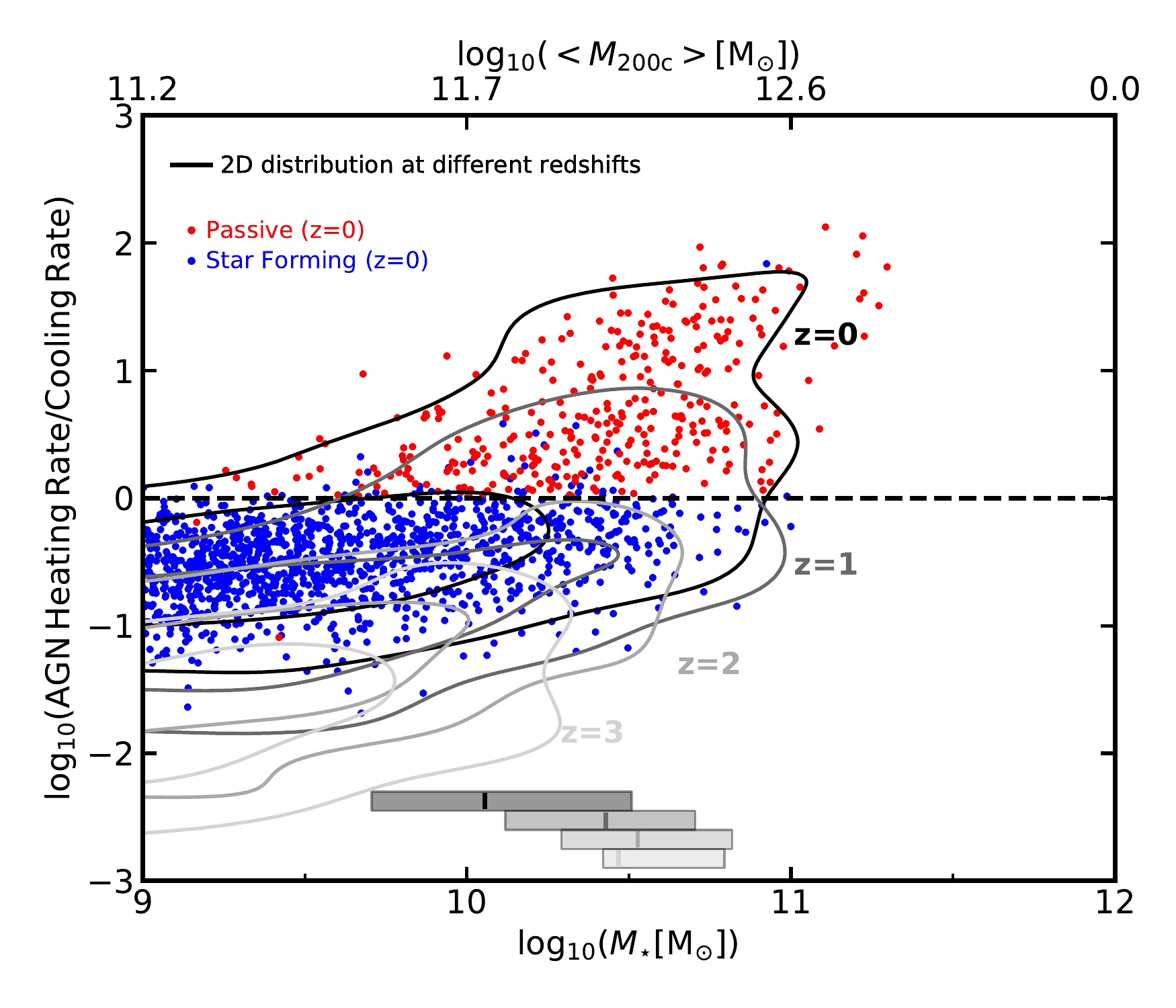}}
\caption{The ratio of maximum AGN radio-mode heating rate to the
  cooling rate from the hot gas atmosphere as a function of stellar
  mass for central galaxies in two different models: AGN feedback 
  only (on the left) and the full H15 model (on the right). The coloured 
  dots show the ratio for individual objects at $z=0$ with red symbols 
  representing quenched galaxies and blue symbols representing 
  star-forming galaxies (divided at $\log_{10}(sSFR[\rm{yr}^{-1}])=-11.0$). 
  The dashed black line divides galaxies where cooling is fully 
  suppressed (above), from galaxies where it still continues at a 
  reduced rate (below). The solid contours on the right panel show 
  the 2D distribution of galaxies at different redshifts, while the vertical 
  lines and solid bars at the bottom of this panel represent the median 
  and the 16\% to 84\% range of stellar mass for galaxies that just 
  crossed the black dotted line upwards (i.e. were just quenched). 
  From top to bottom the bars refer to redshifts z=0, 1, 2 and 3. The top
  $x$-axes show the average $M_{\rm{200c}}$ for a given stellar mass at
  $z=0$. Since galaxies with $\log_{10}(M_*/\Msun) \sim 9$ are hosted 
  by haloes with very low mass in the AGN-only model we only show 
  results based on the MRII simulation for this plot.}
\label{fig:heating_cooling}
\end{figure*}

\subsection{The AGN quenching regime}
\label{subsec:heating_cooling}

The previous subsection detailed how our SN feedback model leads to a
peak in cold gas fraction and hence to enhanced black hole growth at
$\log_{10}(M_\mathrm{200c}/\Msun) \sim 12$ where the proportion of hot gas is
also rapidly increasing. We now explore why this results in AGN
triggering and the quenching of star formation at this same mass
scale. In order to better understand the quenching process, we will
look at the balance between AGN heating and cooling of hot
halo gas as a function of stellar mass. In the H15 model, black holes are assumed
to grow predominately via cold gas accretion during galaxy mergers:
\begin{equation} \label{eq:quasar}
\Delta M_{\rm BH,Q}=\frac{f_{\rm BH}(M_{\rm sat}/M_{\rm
    cen})\,M_{\rm cold}}{1+(V_{\rm{BH}}/V_{\twoc})^2},
\end{equation}
while AGN heating, $\dot E_{\rm{radio}} =\eta \dot M_{\rm{BH}}c^2$, 
is assumed to be proportional to the product of black hole mass and hot gas mass:

\begin{equation} \label{eq:radio}
  \dot{M}_{\rm{BH}}=k_{\rm{AGN}}
  \left(\frac{M_{\rm{hot}}}{10^{11}\Msun}\right)\left(\frac{M_{\rm{BH}}}{10^8\Msun}\right)
\end{equation}
and $f_{\rm BH}$, $V_{\rm{BH}}$ and $k_{\rm{AGN}}$ are tuneable parameters.
 
As halo masses approach $10^{12}\Msun$, hot gas fractions and BH masses 
both grow rapidly, leading to a dramatic increase in feedback from AGN. For the 
parameters selected by our MCMC chains, this results in the 
complete suppression of cooling above a relatively narrow range in
halo masses only slightly above the maximum for which SN
feedback is able to maintain gas outside of haloes. 
We have found that this behaviour is still present in models without the 
$V_{200c}$ dependence in equation~\ref{eq:quasar} and/or without the $M_{\rm{hot}}$
dependence in equation~\ref{eq:radio}. It is important to highlight
that our AGN feedback model only shuts down cooling without ejecting 
any gas from the halo. The subsequent quenching of star formation 
happens due to starvation after all the remaining cold gas is turned into stars (a similar
behavour was inferred from the metal content of observed quenched galaxies
in \citealt{Peng2015}).

Fig.~\ref{fig:heating_cooling} compares the balance between AGN
heating and hot gas cooling as a function of the stellar mass of the
central galaxy for two different models: AGN-only (on the left) and
the full H15 model (on the right). The coloured dots show the values for 
individual galaxies at $z=0$ with red symbols representing quenched 
galaxies and blue symbols representing star-forming objects (divided at
$\log_{10}(sSFR[\rm{yr}^{-1}])=-11.0$). Note that in order to enhance
the legibility of the plot, we here represent the heating rate by the
value from equation~\ref{eq:radio}, whereas in the model it is assumed
to saturate when it equals the cooling rate. The dashed black line
separates galaxies where cooling is fully suppressed (above) from
galaxies where it is still occurring (below). The solid contours 
  on the right panel show the 2D distribution of galaxies at different redshifts, while 
  the vertical lines and solid bars at the bottom of this panel represent the median and 
  the 16\% to 84\% range of stellar mass for galaxies that just crossed the black dotted 
  line upwards (i.e. were just quenched). From top to bottom the bars refer to 
  redshifts z=0, 1, 2 and 3. The top $x$-axis shows
the average $M_{\rm{200c}}$ for each stellar mass at $z=0$. 
Note that this axis differs between the two panels. Since galaxies
with $\log_{10}(M_*/\Msun) \sim 9$ are hosted by haloes with very low
mass in the AGN-only model we only show results based on the Millennium-II
simulation for this plot.

Without SN feedback (left panel) we see that galaxies of a given
stellar mass lie in substantially lower mass haloes than in the H15
model, and that in almost all haloes with $\log_{10}(M_{\rm{200c}}/\Msun) \gtrsim 10.5$
the cooling from the hot gas halo is
fully suppressed.  This is in part because they have, on average, more
massive black holes (which increases the heating rate) and in part
because they have lower amounts of hot gas (which decreases the
cooling rate more than the heating rate). Despite this suppression of
cooling, an excessively large fraction of galaxies with $\log_{10} (M_*/\Msun) <
10.5$ are star-forming in the AGN-only case. This reflects the large
amount of cold gas available (since none is removed by SN feedback)
which leads to more massive black holes and to longer gas exhaustion
times at given stellar mass, and to much larger black hole and cold
gas masses at given halo mass. 

The situation is very different when SN feedback is included (right
panel). The large hot gas ejection rates in low-mass galaxies result in much
lower stellar and gas fractions at given halo mass, and in weak black
hole growth for $\log_{10} (M_*/\Msun)\lesssim10.5$ corresponding to
$\log_{10}(M_{\rm{200c}}/\Msun)\lesssim12$. This produces an almost flat
heating-to-cooling ratio in this regime. As the halo mass for which SN 
ejection becomes ineffective is approached, 
black hole and hot gas masses become large enough that AGN feedback can 
prevent all cooling of hot gas. Since SN feedback restricts the cold gas 
available in all galaxies, the suppression of cooling leads to gas exhaustion and
to a clear separation between star forming and quenched galaxies, respectively 
below and above the cooling dominated/heating dominated transition line.
It is clear from Fig.~\ref{fig:heating_cooling} that
the stellar mass scale at which this transition occurs evolves only
weakly for $z\le 3$ (the solid bars at the bottom of the right panel show significant
overlap). This is a consequence of the shape of the relation between 
heating-to-cooling ratio and stellar mass, in particular, of its change in slope at
$\log_{10}(M_*/\Msun)\sim10.5$ (corresponding to
$\log_{10}(M_{\rm{200c}}/\Msun) \sim 12$).

\begin{figure*}
\centering
\includegraphics[width=1.0\linewidth]{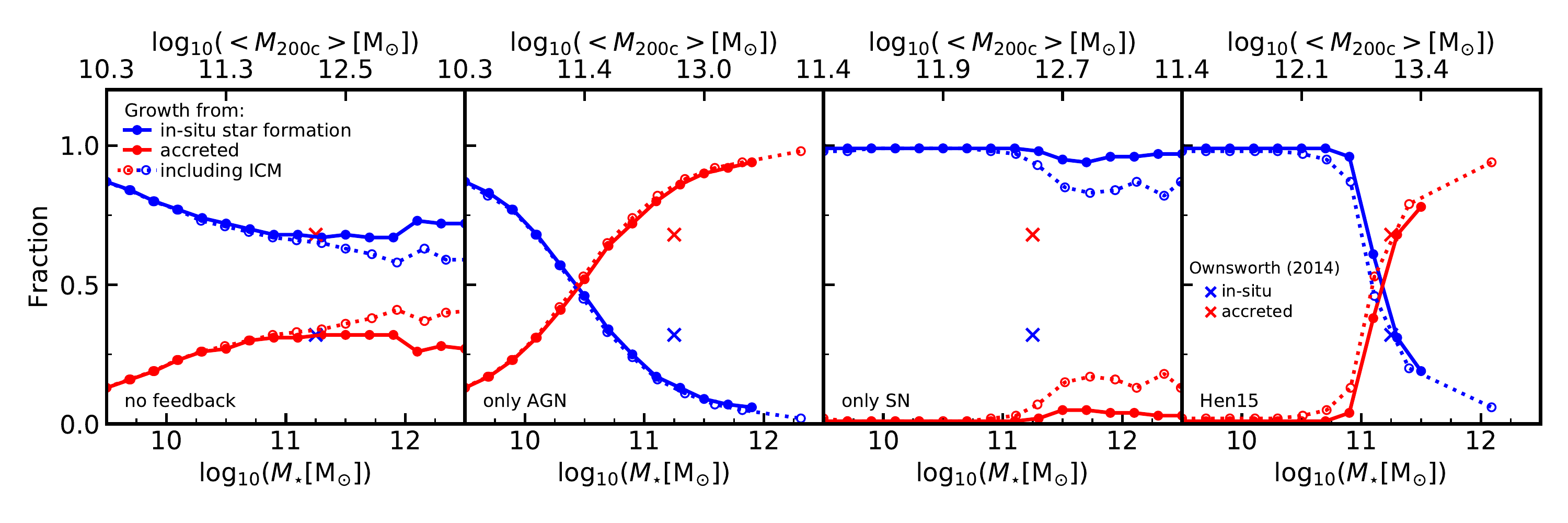}
\caption{Stellar mass fractions at $z=0$ from {\it in situ} star
  formation (solid blue lines) and from merging satellites (solid red
  lines) as a function of stellar mass. The dotted lines show the same quantities but including 
intra-cluster stars. Starbursts are always
  negligible in our model (contributing $\lesssim3\%$) and are not
  shown. From left to right, the panels give results for models with
  no feedback, with AGN feedback only, with SN feedback only, and with
  the full \citet{Henriques2015} feedback model.  The top $x$-axis in
  each panel shows the median halo mass, $M_{200c}$, at given central galaxy stellar
  mass (without intracluster stars). Note that it differs from panel to panel. 
  This plot is restricted to MR-II based results in order to correctly 
  resolve properties that strongly depend on merger histories. Coloured 
  crosses show observational data from \citet{Ownsworth2014}, assuming 
  similar in-situ and accreted fractions below and above $z=3$. } 
\label{fig:channels}
\end{figure*}

\subsection{How important are mergers?}
\label{sec:channels}

Mergers play a major role in regulating feedback in our model, because
black hole growth occurs primarily through merger-driven accretion of
cold interstellar gas. In addition, of course, mergers provide a
second channel for galaxies to increase their stellar mass in addition
to {\it in situ} star formation.  In this section we investigate how
the relative importance of these two channels is affected by the
feedback processes we are investigating.

In Fig.~\ref{fig:channels} we show the stellar fractions at $z=0$
coming from {\it in situ} formation (blue) and from merging satellites
(red). The dotted lines show the same quantities but including 
intra-cluster stars (ICM, hereafter) which originate from tidally disrupted satellites.
We do not separate out stars from merger-driven starbursts
since this channel is negligible in our model ($\lesssim3\%$) 
and restrict our analysis to Millennium-II based results in order to correctly 
resolve properties that strongly depend on merger histories. 
The behaviour in these plots results from the interplay
between the statistics of halo mergers and the halo occupation
distribution of galaxies. While the former has no dependence on
baryonic processes in our model, the latter varies strongly.  Halo
merger trees are almost self-similar over the range of masses relevant
here, with only a very weak dependence of merger rates on halo mass
(approximately $\propto M_{200c}^{0.1}$, see
\citealt{Fakhouri2010}). 

Fig.~\ref{fig:smhm} shows that at low halo mass
the stellar mass fraction is large and approximately constant in
models without stellar feedback (no-feedback and AGN-only 
models, two left panels), while when SN feedback is included 
(SN-only and H15 models, two right panels),
it rises rapidly with halo mass to a peak near $\log_{10}
(M_{200c}/\Msun) \sim 12$. In all models the stellar mass fraction
decreases at high halo mass (for $\log_{10} (M_{200c}/\Msun) \gtrsim$
11.0 and 12.0 in the cases without and with SN feedback,
respectively). The top axes in Fig.~\ref{fig:channels} show that these
halo masses correspond to stellar masses where the fraction of
accreted stars starts to rise substantially ($\log_{10} (M_*/\Msun)=10.0$ 
and 11.0 in the cases without, and with SN feedback, 
respectively two left and two right panels). At lower mass the fraction 
of accreted stars drops slowly when SN feedback is absent and very 
rapidly when it is included, corresponding to the 
different slopes at low mass in Fig.~\ref{fig:smhm}.

In detail, we see that in the absence of feedback (left panel), merger
contributions to the final stellar mass of galaxies increase slowly as
cooling efficiencies drop, but still only amount to $\sim25\%$ at 
$\log_{10}(M_*/\Msun)=12$ ($\sim40\%$ including ICL). For the 
AGN-only model (second panel from the left) the rise in the merger 
contribution at high mass is much stronger, corresponding to the steep 
decline of in-situ star formation caused by the AGN-driven suppression of cooling; massive 
galaxies are almost all quenched by AGN feedback and can only grow 
by mergers. Furthermore, in the absence of SN feedback, lower mass 
satellite halos contain relatively massive stellar components which contribute 
significantly to the central galaxy when they merge.  In the case of the SN-only
model (second panel from the right), the contribution from mergers is
negligible at all masses for the opposite reason; 
regulation of star formation due to SN winds in low mass satellites 
means that they contribute relatively few stars to the more massive centrals 
they merge with. Massive galaxies are almost all star-forming and the {\it in
  situ} contribution is dominant in this model. When both feedback
channels are present, mergers contribute a negligible fraction of the
stars at low mass, for the same reason as in the SN-only model (the stellar
content is reduced in low mass satellite galaxies), and a
significant fraction of the stars at high mass, for the same reason as
in the AGN-only model (quenched galaxies can only grow through mergers).

In our model, stars accreted from mergers invariably end up residing in a 
bulge component. Hence, the red lines in Fig. 7 can be read additionally as the 
fraction of stars in bulges (with the blue lines indicating the fraction of stars in disks). 
Consequently, in the presence of AGN feedback (second and fourth panels of Fig. 7) 
high mass galaxies are bulge dominated structures quenched due to AGN feedback. 
Thus, our model offers an explanation to the strong observational link between 
bulge-to-total ratio and quenching at fixed stellar mass 
\citep[e.g.][]{Bluck2014, Lang2014, Omand2014} via AGN feedback 
(as argued for in \citealt{Bluck2016} and \citealt{Teimoorinia2016})
without requiring quenching to be driven explicitly by morphology as advocated, 
for example, by \citet{Martig2009}.

\begin{figure}
\advance\leftskip-0.45cm
\includegraphics[width=1.1\linewidth]{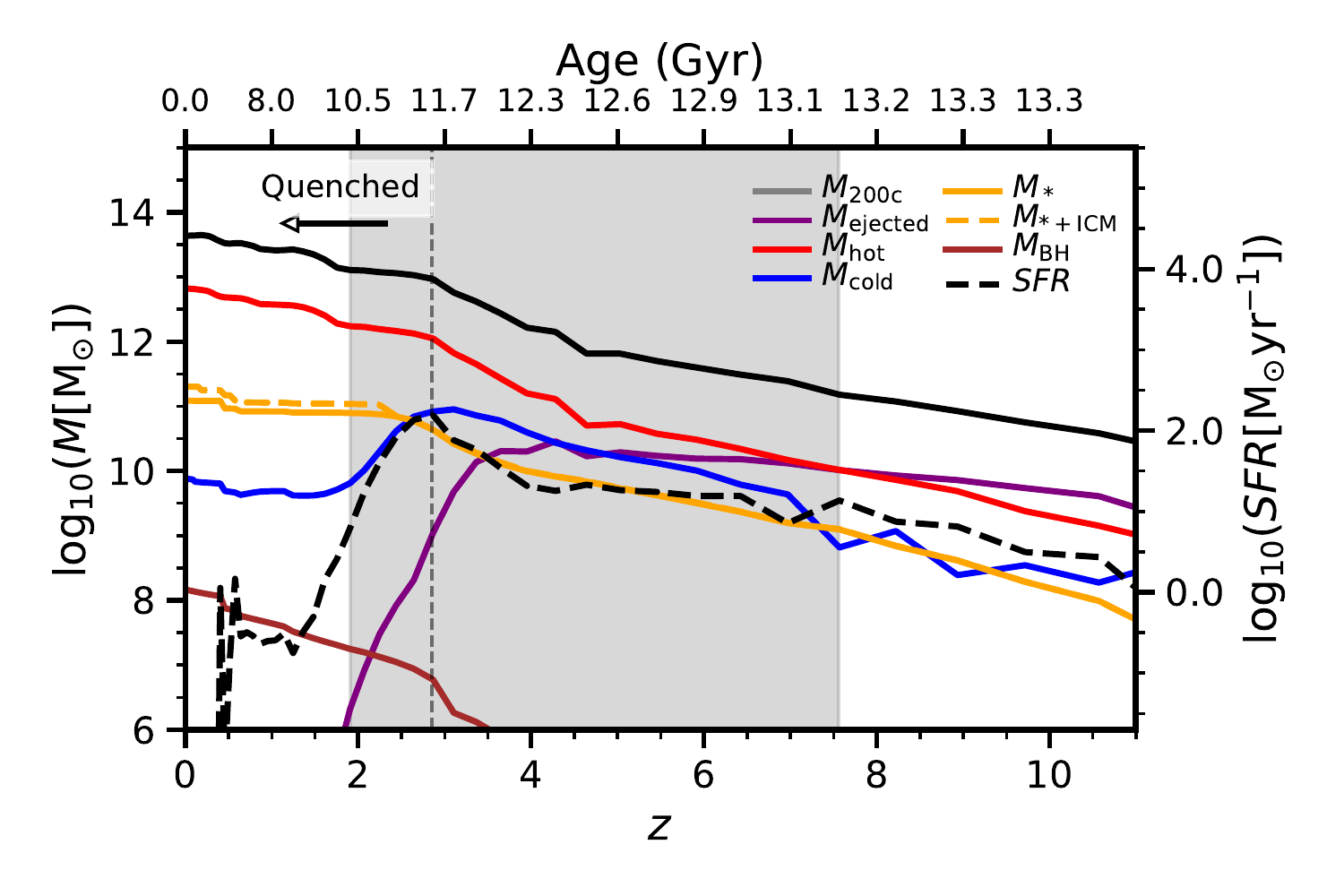}
\caption{The evolution of various properties as functions of redshift
  (bottom $x$-axis) or age (top $x$-axis) for a typical massive
  central galaxy in H15 model ($\log_{10}(M_*/\Msun) \sim11.1$ at $z=0$) that
  quenched at $z\sim2.0$ ($sSFR<(1+z)/2t_{\mathrm{H}}$, 
  following \citealt{Terrazas2016a}). The solid black, purple, blue, red, yellow
  and brown lines represent the evolution of the masses of the halo,
  the ejected gas, the cold gas, the hot gas, the stars and the black
  hole, respectively, with values shown on the left $y$-axis.  The
  dashed yellow line shows the stellar mass including stars assigned
  to the intracluster light component. The dashed black line shows the
  evolution of the SFR with values shown on the right $y$-axis. The shaded 
  region highlights the redshift range between the time when the hot 
  component becomes more massive than the external reservoir and that 
  when star formation is quenched. The vertical dashed lines mark the 
  peak of star formation activity.}
\label{fig:galaxy_evo}
\end{figure}

\section{Discussion}
\label{sec:conclusions}

The complex interplay between the various physical processes included
in our model is perhaps best appreciated by studying the evolution of
a typical massive galaxy.  In Fig.~\ref{fig:galaxy_evo} we show the
growth of the various components of a galaxy which at $z=0$ has
stellar mass $\log_{10} (M_*/\Msun)\sim11.1$ and is the central object
in a group of total mass $\log_{10} (M_{200c}/\Msun)\sim 13.6$.  The
solid black, purple, blue, red, yellow and brown lines represent the
evolution of the masses of the halo, the ejected gas, the cold gas,
the hot gas, the stars and the black hole, respectively, with values
shown on the left $y$-axis. The dashed yellow line shows the stellar
mass including stars assigned to the interstellar light, while the
dashed black line shows the evolution of the SFR with values shown on
the right $y$-axis. The shaded region highlights the redshift range 
between the time when the hot component becomes more massive 
  than the external reservoir and that when star formation is quenched. 
  The vertical dashed lines mark the peak of star formation activity

At $z>7$, the halo is small enough to allow supernova feedback to 
eject large amounts of hot gas and the external reservoir contains 
most of the baryons. After $z\sim 7$, however, ejection and reincorporation 
of gas balance out (high redshift boundary of the shaded region) and the 
mass in the external reservoir remains almost constant until $z\sim 3$ when 
the halo mass rises above $10^{12}\Msun$ and ejection ceases. Since 
the reincorporation time (assumed inversely proportional to $M_{200c}$) 
has by then dropped below the age of the universe, all ejecta are rapidly 
reabsorbed by the hot halo. Over the whole period $0\leq z<7$ the hot 
gas mass increases in step with the halo mass and dominates the total 
baryon budget of the system.

For $z > 7$ the mass in cold gas is similar to that in stars and the SFR is 
a few $\Msun$/yr. However, as SN ejection efficiency decreases and the 
hot component becomes dominant, the growth of the amount of cold gas 
accelerates until it exceeds the stellar mass by a factor of several. This
induces a steep increase in the SFR reaching values of $\sim$ one hundred 
$\Msun$/yr (vertical dashed line). A merger at around $z\sim 3.0$ leads to 
the dumping of $\sim 10^7\Msun$ of cold gas onto the central black hole, 
which is then massive enough for AGN feedback to shut down further 
cooling from the hot gas halo. As a result, the cold gas mass and the SFR 
plummet and the galaxy quenches (low redshift boundary of the shaded region).  
There is very little growth in the stellar mass of the central galaxy after 
quenching, since the H15 model assigns the stars from disrupting and merging 
satellites primarily to the intracluster light component (compare the dashed 
and solid yellow lines in Fig.~\ref{fig:galaxy_evo}).

Comparison with Fig.~\ref{fig:mass_frac_evo} shows that the behaviour
of the galaxy of Fig.~\ref{fig:galaxy_evo} is typical for its halo
mass, and that the same happens to central galaxies of lower mass haloes but
at later times. When the halo mass reaches $\sim10^{12}\Msun$ 
(corresponding to $M_*\sim 10^{10}\Msun$), SN stop being able to eject gas from haloes
and galaxies experience a phase of increased accretion, intense star formation and accelerated black
hole growth. As indicated by the peak in SFR just before quenching in 
Fig.~\ref{fig:galaxy_evo} (dashed black line), this phase is even more dramatic if the galaxy 
experiences a major merger. For the preferred parameters of the H15 model this
occurs at about the same mass at which haloes transition between the
rapid cooling (``cold flow'') and the static atmosphere (``cooling
flow'') regimes. The rapid growth of the central black hole then
coincides with that of a hot gas atmosphere, resulting in greatly
increased AGN feedback. The relatively narrow range in black hole mass at
which cooling is offset and galaxies are quenched 
($M_{\rm{BH}}\sim 10^7\Msun$) is reached shortly 
thereafter, hence also in a narrow and weakly evolving range in halo mass. 

When the observed evolution of galaxy abundances and red fractions as a 
function of stellar mass are used as constraints, our MCMC sampling of 
parameter space requires (1) a strong inverse relation between the ability
 of SN to eject gas from the halo and $V_{\rm{max}}$ as well as (2) a direct 
 relation between reincorporation of this gas and $M_{200c}$. In the hydro 
simulations of \citet{Dubois2015} and \citet{Bower2017} this behaviour seems to 
result from the inability of SN winds to penetrate a static hot atmosphere. 
Independently of the specific modelling implementation, a strong dependence of SN 
ejection efficiency on halo mass, coupled with an AGN feedback model grounded 
on black hole growth via cold gas accretion, can easily result in roughly constant 
characteristic halo and stellar masses for quenching and maximal conversion of 
baryons into stars. Particularly interesting is the presence of this behaviour in a 
model where quenching is caused by AGN feedback 
and not directly linked to halo or stellar mass.

\section{Summary}
\label{sec:summary}

The present work aims at understanding the physical origin behind two 
observational trends: the redshift independence of the characteristic halo and 
stellar masses at which galaxies quench; and their coincidence with the scale at 
which baryon conversion into stars maximises.

The same behaviour is seen in the H15 model, where it can be 
understood as a consequence of galaxies passing through the following three 
evolutionary phases:
\begin{itemize}
  \item the stellar and halo masses of central galaxies reach a roughly 
  redshift-independent mass at which supernovae are no longer able to eject 
  material to large radii where it can remain unavailable for reaccretion;  
  \item subsequently, enhanced accretion of hot halo gas onto galaxies produces 
  a relatively large cold gas to halo mass ratio, causing enhanced star 
  formation and enhanced black hole growth 
  \item soon thereafter, the masses of the black hole and of the hot gas halo have grown to the point 
  where AGN feedback stops all further cooling onto the galaxy, leading to quenching as it runs out of 
  fuel, and leaving it with a stellar to halo mass ratio which is close to maximal.
\end{itemize}

Within such a model, the specific values selected for
the 17 parameters allowed to vary in the MCMC chains of H15 are required 
to obtain consistency with the observed abundance and
passive fraction of galaxies as a function of stellar mass over the
range $0\leq z \leq 3$. As Fig.~S9 of H15 shows, the values of all 17
parameters are quite well constrained, so the coincidences and
regularities we have focused on in this paper can all be viewed as
required by the observational data, at least in the context of models
of this type. The actual values obtained for the parameters can then
be taken as observational estimates for the efficiencies and the
scaling properties of the various cooling, star formation,  black hole
formation and feedback processes considered. More detailed simulation
work is, of course, required to check that the representation of these
processes is appropriate, and to understand {\it why} the efficiencies
and scalings have the values the observations appear to require.

\section*{Acknowledgements}

BMBH (ORCID 0000-0002-1392-489X) acknowledges support from a Zwicky Prize fellowship.
B.A.T. is supported by the National Science Foundation
Graduate Research Fellowship under Grant No. DGE
1256260.

\bibliographystyle{mn2e} \bibliography{paper_HWL18}

\label{lastpage}

\end{document}